\documentclass[aps,pra,groupedaddress,showpacs,twocolumn]{revtex4}%
\usepackage{amsfonts}
\usepackage{amsmath}
\usepackage{amssymb}
\usepackage{graphicx}%
\begin{document}
\preprint{HEP/123-qed}
\title{Bound state and non-Markovian dynamics of a quantum emitter around a surface plasmonic nanostructure}
\author{Sha-Sha Wen}
\affiliation{College of Physics, Mechanical and Electrical Engineering, Jishou University,
Jishou 416000, China }
\author{Yong-Gang Huang}
\email{huang122012@163.com}
\affiliation{College of Physics, Mechanical and Electrical Engineering, Jishou University,
Jishou 416000, China }
\author{Xiao-Yun Wang}
\email{wxyyun@163.com}
\affiliation{College of Physics, Mechanical and Electrical Engineering, Jishou University,
Jishou 416000, China }
\author{Jie Liu}
\affiliation{College of Physics, Mechanical and Electrical Engineering, Jishou University,
Jishou 416000, China }
\author{Yun Li}
\affiliation{College of Physics, Mechanical and Electrical Engineering, Jishou University,
Jishou 416000, China }
\author{Ke Deng}
\affiliation{College of Physics, Mechanical and Electrical Engineering, Jishou University,
Jishou 416000, China }
\author{Xiu-E Quan}
\affiliation{College of Physics, Mechanical and Electrical Engineering, Jishou University,
Jishou 416000, China }
\author{Hong Yang}
\affiliation{College of Physics, Mechanical and Electrical Engineering, Jishou University,
Jishou 416000, China }
\author{Jin-Zhang Peng}
\affiliation{College of Physics, Mechanical and Electrical Engineering, Jishou University,
Jishou 416000, China }
\author{He-Ping Zhao}
\affiliation{College of Physics, Mechanical and Electrical Engineering, Jishou University,
Jishou 416000, China }
\pacs{22}

\begin{abstract}
A bound state between a quantum emitter (QE) and surface plasmon polaritons (SPPs) can be formed, where the QE is partially stabilized in its excited state.  We put forward a general approach for calculating the energy level shift at a negative frequency $\omega$, which is just the negative of the nonresonant part for the energy level shift at positive frequency $-\omega$. We also propose an efficient formalism for obtaining the long-time value of the excited-state population without calculating the eigenfrequency of the bound state or performing a time evolution of the system, in which the probability amplitude for the excited state in the steady limit is equal to one minus the integral of the evolution spectrum over the positive frequency range. With the above two quantities obtained, we show that the non-Markovian decay dynamics in the presence of a bound state can be obtained by the method based on the Green's function expression for the evolution operator. A general criterion for identifying the existence of a bound state is presented. These are numerically demonstrated for a QE located around a nanosphere and in a gap plasmonic nanocavity. These findings are instructive in the fields of coherent light-matter interactions.
\end{abstract}
\volumeyear{year}
\volumenumber{number}
\issuenumber{number}
\eid{identifier}
\date[Date text]{date}
\received[Received text]{date}

\revised[Revised text]{date}

\accepted[Accepted text]{date}

\published[Published text]{date}

\startpage{101}
\endpage{102}
\maketitle

\section{ Introduction}
Coherent interaction between a quantum emitter (QE), such as atom, molecule or
quantum dot, and the quantized electromagnetic field lies at the heart of
quantum optics
\cite{cohen1989photons,agarwal1974quantum,scully1997quantum,peter1994quantum,PhysRevLett.118.237401,PhysRevLett.118.073604,Zhang2017Sub,santhosh2016vacuum}%
. An incredibly wide observable phenomena, such as ultrafast single-photon
switches \cite{volz2012ultrafast,chen2013all,tiecke2014nanophotonic}, trapping atoms by vacuum forces
\cite{chang2014trapping,haroche1991trapping,cook1982electromagnetic,Zhang:18},
photon blockade \cite{birnbaum2005photon,ridolfo2012photon}, single-molecule sensing \cite{miles2013single}, single-atom lasers
\cite{mckeever2003experimental,oulton2009plasmon}, enhanced and inhibited
spontaneous emission
\cite{PhysRevLett.55.2137,PhysRevLett.58.2486,lodahl2004controlling,Xue2003Decay,PhysRevLett.100.203002,Noda2007Spontaneous,PhysRevB.89.041402,PhysRevLett.112.253601}%
, giant Lamb shift \cite{PhysRevLett.84.2136,wang2004giant}, quantum nonlinear optics \cite{peyronel2012quantum,chang2014quantum}, bound state
\cite{PhysRevLett.64.2418,PhysRevX.6.021027,PhysRevLett.121.220403,PhysRevA.87.052139,PhysRevA.50.1764,PhysRevE.93.012107,PhysRevA.93.033833,Tong_2010,liu2017quantum,gaveau1995limited,PhysRevA.81.052330,wu2014quenched,PhysRevLett.104.023602,PhysRevA.93.033829,PhysRevA.93.013828,PhysRevLett.113.263604,PhysRevA.87.022312,PhysRevE.90.022122,PhysRevA.94.043839,PhysRevLett.109.170402}
, etc, have been predicted and demonstrated. Among all, bound state is one particular example, where a photonic excitation is confined to the vicinity of the QE and a discrete eigenstate is formed inside the environmental photonic band gap. The quantum coherence, which plays a prominent role in the fields of quantum  physics and technology \cite{steane1998quantum,saffman2010quantum,Buluta_2011,RevModPhys.89.035002,you2011atomic,PhysRevA.93.020105,Wang_2017,PhysRevLett.58.2059}, can be protected from environmental noise.

Recently, a bound state between a QE and surface plasmon polaritons (SPPs) is predicted \cite{PhysRevB.95.161408}, where the QE does not decay completely to its ground state and part of its excited-state population is trapped in the steady state even in the presence of the lossy metal. Different from the previous investigation \cite{PhysRevLett.64.2418,PhysRevX.6.021027,PhysRevLett.121.220403,PhysRevA.87.052139,PhysRevA.50.1764,PhysRevE.93.012107,PhysRevA.93.033833,Tong_2010,liu2017quantum,gaveau1995limited,PhysRevA.81.052330,wu2014quenched,PhysRevLett.104.023602,PhysRevA.93.033829,PhysRevA.93.013828,PhysRevLett.113.263604}, it is an open quantum system and there is no photonic band gap in the positive frequency domain. The formation of a QE-SPP bound state has been attributed to the strong field confinement which can greatly enhance their interaction. Inspired by the above works, we are interested in the conditions for the existence of a bound state for a QE located around a plasmonic nanostructre.

To determine the existence or absence of a bound state, one usually calculates the long-time values of the excited-state population by solving the non-Markovian master equation, in which a time-convolution integral should be performed. As discussed in Ref. \cite{PhysRevA.99.053844}, this method is time consuming. Alternatively,  one can determine the existence or absence of a bound state by the resolvent operator method \cite{cohen1989photons,agarwal1974quantum,PhysRevA.85.053827,garmon2019non,bay1997fluorescence,PhysRevA.99.053844,PhysRevLett.109.170402}, once the energy level shift (Lamb shift) is known. Very recently, we have proposed a general numerical method for calculating the energy level shift of a QE in an arbitrary nanostructure for positive frequency \cite{PhysRevA.99.053844}. In this work, we will generalize the above method to the negative frequency domain, in which the eigenfrequency for a bound state is located. In addition, we will propose a general criterion for identifying the existence of a bound state for a QE located in an arbitrary nonostructure. We will show that the photon Green's function (GF) at zero frequency, which can be calculated by numerous methods \cite{Zhao:18,tai1994dyadic,Bai:13,doi:10.1002/lpor.201500122,PhysRevX.5.021008,VanVlack:12,lalanne2018light,Tian_2019,yunjinwulixuebao,hohenester2012mnpbem,jackson2007classical,novotny2012principles},  is suffice. This can greatly simplify the problem. Besides, we provide two different formalisms to easily determine the excited-state population in the steady state without performing a temporal evolution.

As demonstrated in Ref. \cite{PhysRevA.99.053844}, the exact non-Markovian dynamics can be investigated by the method based on the Green's function expression for the evolution operator, in which there is no need of time convolution. For arbitrary nanostructure, one just needs to calculate the photon GF in the real frequency domain, in which no particular assumption about the  permittivity of the material, such as the Drude model, should be made. It removes most limitations, such as the usual tight-binding or a quadratic dispersion assumption for the case of a cavity array or photonic crystal, encountered in previous analytical approaches and allows us to solve the exact decay dynamics for an open system in an almost fully numerical way. In this work, we would like to generalize the above method to study the decay dynamics of a quantum emitter coupled to surface plasmons when a bound state is formed.

The remainder of this paper is organized as follows. In Sec. II, we first present the theory and derive a general method for calculating the energy level shift in the negative frequency domain. Besides, we then provide a general criterion for identifying the existence of a bound state and two new formalisms for calculating the excited-state population in the steady state. In Sec. III, we apply our method to a particular example where a QE is located around a single gold nanosphere. The performance of the above methods for energy level shift in negative frequency domain, condition for the existence of bound state, excited-state population in the steady state and non-Markovian dynamics are shown. Section IV is devoted to the study of the bound state and non-Markovian dynamics of a QE located in a plasmonic nanocavity. Finally, we conclude in Sec. V.

\section{Theory and Method}

Under the dipole and rotating-wave approximations, the total Hamiltonian for a
two-level QE coupled to a common electromagnetic reservoir is
\cite{PhysRevA.68.043816}
\begin{align*}
H  &  =H_{0}+H_{I},\\
H_{0}  &  =\int d\mathbf{r}\int_{0}^{+\infty}d\omega\hbar\omega\text{ }%
\hat{\mathbf{f}}^{\dagger}\left(  \mathbf{r},\omega\right)  \cdot
\hat{\mathbf{f}}(\mathbf{r},\omega)+\hbar\omega_{0}|e_{0}\rangle\langle
e_{0}|,\\
H_{I}  &  =-\int_{0}^{+\infty}d\omega\lbrack|e\rangle\langle g|\mathbf{d}%
^{\ast}\cdot\hat{\mathbf{E}}\left(  \mathbf{r}_{0},\omega\right)
+\mathrm{H.c}.].
\end{align*}
Here, $\hat{\mathbf{f}}\left(  \mathbf{r},\omega\right)  $ and $\hat
{\mathbf{f}}^{\dagger}\left(  \mathbf{r},\omega\right)  $ are the annihilation
and creation operators for the elementary excitation of the electromagnetic
reservoir, respectively. $\mathbf{d}=\langle g|\hat{\mathbf{d}}|e\rangle
=d\hat{\mathbf{n}}$ is the matrix element for the transition dipole moment,
with the unit vector $\hat{\mathbf{n}}$ and its strength $d$. The electric
field vector operator $\hat{\mathbf{E}}\left(  \mathbf{r},\omega\right)  $ is
given by $\hat{\mathbf{E}}(\mathbf{r},\omega)=i\sqrt{\frac{\hbar}%
{\pi\varepsilon_{0}}}\int d\mathbf{s}\sqrt{\varepsilon_{I}(\mathbf{s},\omega
)}\mathbf{G}(\mathbf{r},\mathbf{s},\omega)\cdot\hat{\mathbf{f}}(\mathbf{s}%
,\omega)$, where $\mathbf{G}(\mathbf{r},\mathbf{s},\omega)$ is the photon GF
defined as [$\nabla\times\nabla\times-\varepsilon(\mathbf{r},\omega)\omega
^{2}/c^{2}]\mathbf{G}(\mathbf{r},\mathbf{s};\omega)=\omega^{2}/c^{2}%
\mathbf{I}\delta(\mathbf{r}-\mathbf{s})$. Here $\varepsilon(\mathbf{r}%
,\omega)=\varepsilon_{R}(\mathbf{r},\omega)+i\varepsilon_{I}(\mathbf{r}%
,\omega)$ is the frequency-dependent complex relative dielectric function in
space and $\varepsilon_{I}(\mathbf{r},\omega)$ is its imaginary part.
$\mathbf{I}$ is the unit dyad and $c$ refers to the speed of light in vacuum.

We assume initially the field is in the vacuum state and the QE is excited. In
this case, the states of interest are $|I\rangle=|e\rangle\otimes|0\rangle$
and $|F_{r,\omega}\rangle=|g\rangle\otimes|1_{r,\omega}\rangle$ with
$|e\rangle$ ($|g\rangle$) the excited (ground) state of the QE and
$|1_{r,\omega}\rangle\equiv\hat{\mathbf{f}}_{j}^{\dagger}\left(
\mathbf{r},\omega\right)  |0\rangle$. $|0\rangle$ is the zero photon state.
The state vector of the system evolves as
\begin{align*}
|\Psi(t)\rangle & \equiv U(t)|I\rangle  \\
& =c_{1}(t)|I\rangle+ \int
dr\int_{0}^{+\infty}d\omega C(r,\omega,t)|F_{r,\omega}\rangle,
\end{align*}
with the initial condition $c_{1}(0)=1$.

As demonstrated in Ref. \cite{PhysRevA.99.053844}, the dynamics can be efficiently
obtained by the resolvent operator technique, in which the probability amplitude for the excited state is  $c_{1}(t)=\langle I|U(t)|I\rangle$ with $U(t)=\int
_{-\infty}^{+\infty}d\omega\lbrack G^{-}(\omega)-G^{+}(\omega)]exp(-i\omega
t)/(2\pi i)$. Here, the retarded and advanced Green's functions are $G^{\pm
}(\omega)=\lim_{\eta\rightarrow0^{+}}G(\omega\pm i\eta)$ with $G(z)=(z-H/\hbar
)^{-1}$. From the operator identity $(z-H_{0}/\hbar)G(z)=H_{I}G(z)/\hbar$, we
obtain $G_{ii}(\omega)\equiv\langle I|G(\omega)|I\rangle=[\omega-\omega
_{0}-R_{ii}(\omega)]^{-1}$, in which the matrix element for the level shift
operator $R_{ii}(\omega)$ reads $R_{ii}(z)=\frac{1}{\pi\varepsilon_{0}}%
[\int_{0}^{\infty}d\omega\frac{\mathbf{d}^{\ast}\cdot\operatorname{Im}%
\mathbf{G}(\mathbf{r}_{A},\mathbf{r}_{A},\omega)\cdot\mathbf{d}}{z-\omega}$.
By using the relation $\lim_{\eta\rightarrow0^{+}}1/(z-\omega-i\eta
)=\mathbb{P}[1/(z-\omega)]+i\pi\delta(z-\omega)$ with $\mathbb{P}$
representing the Cauchy principal value, one has $R_{ii}^{\pm}(z)=\lim
_{\eta\rightarrow0^{+}}R_{ii}(z\pm i\eta)=\Delta(z)\mp i\frac{\Gamma(z)}{2}$,
in which the spontaneous emission rate $\Gamma(z)$ and the energy level shift
$\Delta(z)$ are%
\begin{align}
\Gamma(z)  &  =2\pi\operatorname{Im}g_{rr}(z)\theta(z),\label{gama}\\
\Delta(z)  &  =\mathbb{P}\int_{0}^{+\infty}ds\frac{\operatorname{Im}g_{rr}%
(s)}{z-s}. \label{delta}%
\end{align}
Here, $\theta(z)$ is the step function and the coupling strength is%
\begin{equation}
g_{rr}(\omega)=\frac{\mathbf{d}^{\ast}\cdot\mathbf{G}(\mathbf{\mathbf{r}_{0}%
},\mathbf{r}_{0},\omega)\cdot\mathbf{d}}{\hbar\pi\varepsilon_{0}}.
\label{couplingstrg}%
\end{equation}

Thus, the probability amplitude for the excited state becomes
\begin{equation}
c_{1}(t)=\int_{-\infty}^{+\infty}S(\omega)e^{-i\omega t}d\omega,
\label{Fourier-Laplacedynamics}%
\end{equation}
with the evolution spectrum
\begin{equation}
S(\omega)=\frac{1}{\pi}\lim_{\eta\rightarrow0_{+}}\frac{\Gamma(\omega)/2+\eta
}{[\omega-\omega_{0}-\Delta(\omega)]^{2}+[\Gamma(\omega)/2+\eta]^{2}}.
\label{spectrum}%
\end{equation}

To evaluate Eq. (\ref{Fourier-Laplacedynamics}), one should calculate the
evolution spectrum $S(\omega)$ as well as the energy level shift
$\Delta(\omega)$ in the whole frequency range. For $\omega\geq0$, we have
demonstrated in Ref. \cite{PhysRevA.99.053844} that the energy level shift can be calculated by the subtractive KK method efficiently. For the sake of
completeness, we include the derivation of the method here. By using the
relations $-\pi\operatorname{Re}\mathbf{G}(\mathbf{\mathbf{r}_{0}}%
,\mathbf{r}_{0},s)=\mathbb{P}\int_{-\infty}^{+\infty}ds\operatorname{Im}%
\mathbf{G}(\mathbf{\mathbf{r}_{0}},\mathbf{r}_{0},s)/(\omega-s)$ and
$\operatorname{Im}\mathbf{G}(\mathbf{\mathbf{r}_{0}},\mathbf{r}_{0}%
,-s)=-\operatorname{Im}\mathbf{G}(\mathbf{\mathbf{r}_{0}},\mathbf{r}_{0},s)$ ,
the energy level shift [Eq. (\ref{delta})] for $\omega\geq0$ can be written as
\cite{PhysRevB.84.075419}
\begin{equation}
\Delta(\omega)=-\pi\operatorname{Re}g_{rr}(\omega)+\mathbb{P}\int_{0}%
^{+\infty}ds\frac{\operatorname{Im}g_{rr}(s)}{\omega+s}. \label{delta0}%
\end{equation}

Subtracting $\Delta(0)$ from $\Delta(\omega)$, we have for $\omega\geq0$%
\begin{align}
\Delta(\omega)  &  =-\pi\operatorname{Re}g_{rr}(\omega)+\Delta c(\omega
),\label{SubtractiveKK}\\
\Delta c(\omega)  &  =\frac{\pi}{2}\operatorname{Re}g_{rr}(0)-\omega\int
_{0}^{+\infty}ds\frac{\operatorname{Im}g_{rr}(s)}{(\omega+s)s}. \label{deltac}%
\end{align}
Here, we have used the relation $\Delta(0)=-0.5\pi\operatorname{Re}g_{rr}(0)$,
since the second term on the right hand side in Eq. (\ref{delta0}) is
$-\Delta(0)$ [see Eq. (\ref{delta})]. The first term $-\pi
\operatorname{Re}g_{rr}(\omega)$ is the resonant part arising from the residua
at the poles and $\Delta c(\omega)$ is the correction part which represents
the nonresonant part of the dispersion potential. As discussed in Ref. \cite{PhysRevA.99.053844},
this method is useful and simplifies the numerical integrals in calculating
the energy level shift, sicne there is no need to worry about the principal
value and the calculation of the GF with imaginary frequency argument. In addition, it converges much more quickly than the method shown in Eq. (\ref{delta}).

To generalize the above method to the case for a negative frequency $\omega<0$, we alternatively use
the formalism of Eq. (\ref{delta}) and subtract $\Delta(0)$ from $\Delta(\omega)$ in a similar way. We have for $\omega<0$
\begin{equation}
\Delta(\omega)=-\frac{\pi}{2}\operatorname{Re}g_{rr}(0)+\omega\int
_{0}^{+\infty}ds\frac{\operatorname{Im}g_{rr}(s)}{(\omega-s)s}=-\Delta
c(-\omega). \label{deltanefre}%
\end{equation}

Since the integrand in Eq. (\ref{deltanefre}) decays faster than that in Eq. (\ref{delta}) for large frequency $s$, it will converge much more quickly than the method shown in Eq. (\ref{delta}). This will be numerically demonstrated in the next section.

Equations (\ref{SubtractiveKK}) and (\ref{deltanefre}) are the main results of our methods for calculating the
energy level shift of a quantum emitter for $\omega\geq0$ and $\omega<0$,
respectively. It is general and does not imply any specific configuration or
system. It should be noted that the energy level shift $\Delta(\omega)$ in the
negetive frequecny range ($\omega<0$) is just the negative of the nonresonant
part of the energy level shift for positive frequency $-\omega$, i.e. $-\Delta c(-\omega)$. Thus, $\Delta(\omega)$ in the
negetive frequecny range ($\omega<0$) can be obtained directly once the
nonresonant part $\Delta c(\omega)$ is obtained for a positive frequency $\omega\geq0$. In
addition, it is a monotonically decreasing function and approaching zero as
$\omega\rightarrow-\infty$, which can be clearly seen from Eq. (\ref{delta}).

With the energy level shift $\Delta(\omega)$ calculated, the evolution spectrum
$S(\omega)$ [Eq. (\ref{spectrum})] can be evaluated. For nonzero
$\Gamma(\omega)$, $S(\omega)$ is of a generalized Lorentzian form
$S(\omega)=\frac{1}{\pi}\frac{\Gamma(\omega)/2}{[\omega-\omega_{0}%
-\Delta(\omega)]^{2}+[\Gamma(\omega)/2]^{2}}$. But for frequency
inside the photonic band gap where $\Gamma(\omega)=0$, the evolution spectrum becomes $S(\omega)=\frac{1}{\pi}\lim
_{\eta\rightarrow0_{+}}\frac{\eta}{[\omega-\omega_{0}-\Delta(\omega)]^{2}%
+\eta^{2}}$. In this case, the evolution spectrum $S(\omega)$ is either zero
or a delta function depending on the zero of the function
\begin{equation}
f(\omega)=\omega-\omega_{0}-\Delta(\omega). \label{rootequation}
\end{equation}

If the solution $\omega_{b}$ is not inside the photonic band gap, $S(\omega)=0$. But
for $\omega_{b}$ inside \cite{PhysRevLett.64.2418,PhysRevX.6.021027,PhysRevLett.121.220403,PhysRevA.87.052139,PhysRevA.99.010102,PhysRevA.81.052330,PhysRevA.87.022312,PhysRevE.90.022122,PhysRevA.93.033833,PhysRevA.93.020105,PhysRevA.94.043839,PhysRevA.91.022328,behzadi2018requirement}, the evolution spectrum becomes a delta function for frequency around $\omega_{b}$, i.e. $S(\omega)=\frac{1}{\pi}\lim_{\eta
\rightarrow0_{+}}\frac{\eta}{[\omega-\omega_{0}-\Delta(\omega)]^{2}+\eta^{2}%
}=Z\delta(\omega-\omega_{b})$, in which $Z$ is a real quantity and can be
written as
\begin{equation}
Z=[1-\Delta^{^{\prime}}(\omega_{b})]^{-1}=[1+\int_{0}^{+\infty}\frac
{\operatorname{Im}g(s)}{(\omega_{b}-s)^{2}}ds]^{-1}. \label{zequation}%
\end{equation}

Different from the previous investigations where the energy level shift $\Delta(\omega)$ and accordingly $Z$ are usually analytically evaluated under a tight-binding or a quadratic dispersion assumption for the case of a cavity array or a photonic crystal, we numerically calculate both by the above methods for an open system in this work. Since there is no photonic band gap in the positive frequency range, the existence of a bound state requires that the eigenfrequency $\omega_{b}$ less than or equal to zero, i.e. $\omega_{b}\leq0$. Furthermore, there is at most one root $\omega_{b}$ for equation $f(\omega)=0$ in the range $\omega\leq0$, since $f(\omega)=\omega-\omega_{0}-\Delta(\omega)$ is a
monotonically increasing function [see Eq. (\ref{delta})]. Thus, when a bound state presents, the probability amplitude in Eq. (\ref{Fourier-Laplacedynamics}) can be written as
\begin{equation}
c_{1}(t)=Ze^{-i\omega_{b}t}+\int_{0}^{+\infty}S(\omega
)e^{-i\omega t}d\omega, \label{dynamics}%
\end{equation}
in which the second term tends to zero due to out-of-phase interference. The first term indicates a bound state, since $\lim
_{t\rightarrow\infty}c_{1}(t)=Ze^{-i\omega_{b}t}$. This indicates that the quantity $Z$ is the amplitude for the excited state in the steady limit.

If one is only interested in the long-time value of the excited-state population
$P_{a}(\infty)=\left\vert c_{1}(\infty)\right\vert ^{2}=Z^{2}$, Eq.
(\ref{dynamics}) at $t=0$ becomes
\begin{equation}
Z=c_{1}(0)-\int_{0}^{+\infty}S(\omega)d\omega. \label{Ztime0}%
\end{equation}

Equation (\ref{zequation}) and (\ref{Ztime0}) are the main results of our
methods for calculating the long-time values of the excited-state population
$P_{a}(\infty)$ without performing a time evolution. One just needs to calculate the photon GF in the real frequency range, which can be used to obtain the energy level shift $\Delta(\omega)$ and the spontaneous emission rate $\Gamma(\omega)$. For the method shown in Eq. (\ref{zequation}), we first solve the transcendental equation
$f(\omega)=\omega-\omega_{0}-\Delta(\omega)=0$ [Eq. (\ref{rootequation})] to find the lowest root $\omega_{b}$ and then perform a general integral [Eq. (\ref{zequation})] to obtain $Z$. But for the method shown in Eq. (\ref{Ztime0}), one just needs to perform a general integral about the evolution spectrum $S(\omega)$ [Eq. (\ref{spectrum})] in the positive frequency range without calculating the root $\omega_{b}$. In the next section, we will demonstrate the performances of both methods [Eq. (\ref{zequation}) and (\ref{Ztime0})].

As stated above, the condition for the appearance of a unique bound state is $\omega_{b}\leq0$. This requires $f(0)\geq0$, since the monotonically increasing function $f(\omega)$ in the negative frequency range approaches $-\infty$ as $\omega\rightarrow-\infty$. Explicitly, this can be written as
\begin{equation}
\Delta(0)=-0.5\pi\operatorname{Re}g_{rr}(0)\leq-\omega_{0},
\label{boundstatetecondition}%
\end{equation}
where we have used the relation $\Delta(0)=-0.5\pi\operatorname{Re}g_{rr}(0)$.

Equation (\ref{boundstatetecondition}) is a general criterion for identifying
the existence of a bound state when there is no photonic band gap for the
electromagnetic reservoir in the positive frequency range. It is only at zero
frequency that one can judge whether a bound state exists or not from the real part for the coupling strength $\operatorname{Re}
g_{rr}(0)=\operatorname{Re}[\mathbf{d}^{\ast}\cdot\mathbf{G}%
(\mathbf{\mathbf{r}_{0}},\mathbf{r}_{0},0)\cdot\mathbf{d}]/\hbar\pi
\varepsilon_{0}$. Here, the photon GF $\mathbf{G}(\mathbf{\mathbf{r}_{0}},\mathbf{r}_{0},0)$ can be obtained by numerous methods \cite{Zhao:18,li1994electromagnetic,tai1994dyadic,Bai:13,doi:10.1002/lpor.201500122,PhysRevX.5.021008,VanVlack:12,lalanne2018light,Tian_2019,yunjinwulixuebao,hohenester2012mnpbem,jackson2007classical,novotny2012principles}.  For example, direct
electrostatics methods based on solving the Poisson equation, such as the method of images, the method based on separation of variables, the finite element method (FEM),
the boundary element method and so on, can be applied. Besides, other methods by extrapolating the solution of the full wave equation near zero frequency can be used [see Ref. \cite{PhysRevA.99.053844,Zhao:18} and references therein]. In this work, we follow the method presented in Ref.
 \cite{PhysRevA.99.053844}, in which the analytical method for a nanosphere and the FEM for an arbitrary-shaped nanostructure are adopted.

Before proceeding further, let us give a brief summary about the theory and methods.  Based on the photon GF formalism, one can expresses the medium-assisted quantized electromagnetic field by the fundamental bosonic vector field and study the exact quantum decay dynamics of a QE in an open quantum system by the resolvent operator method. For any nanostructure, the photon GF can be obtained by the method shown in Ref. \cite{PhysRevA.99.053844,Zhao:18}. Then, the coupling strength $g_{rr}(\omega)$ and accordingly the spontaneous emission rate $\Gamma(\omega)$ can be obtained directly by Eq. (\ref{couplingstrg}) and (\ref{gama}), respectively. The energy level shift $\Delta(\omega)$ can be calculated by Eq. (\ref{SubtractiveKK}) for positive frequency ($\omega\ge0$), in which the correction part $\Delta c(\omega)$ can be calculated by Eq. (\ref{deltac}). For $\omega\leq0$, $\Delta(\omega)$ can be obtained directly from Eq. (\ref{delta}) or Eq. (\ref{deltanefre}) with $\Delta c(-\omega)$ calculated by Eq. (\ref{deltac}) for a positive frequency $-\omega$. The  non-Markovian decay dynamics can be calculated by Eq. (\ref{dynamics}). In this equation, we set $Z=0$ if there is no bound state. Otherwise, $Z$ can be obtained by Eq. (\ref{zequation}) or Eq. (\ref{Ztime0}). Both can be used to determine the long-time value of the excited-state population $P_e(\infty) = |c_1(\infty)|^2= Z^2$. It should be noted that there is no need to evaluate the bound state eigenfrequency $\omega_{b}$ through Eq. (\ref{Ztime0}) as compared with the method by Eq. (\ref{zequation}). The bound state eigenfrequency $\omega_{b}$ can be obtained by solving the transcendental equation
$f(\omega)=\omega-\omega_{0}-\Delta(\omega)=0$ [Eq. (\ref{rootequation})]. We have provided a general and easy way to judge whether a bound state exists or not [Eq. (\ref{boundstatetecondition})].

In the following sections, we will first demonstrate the performances of the above methods. For this purpose, we choose a particular example where a two-level QE is located above a metal nanosphere. As shown in Fig. 1(a), a nanosphere with radius $a$ is located at the origin. A QE at a distance $h$ from the surface of the
sphere lies on the z-axis. The metal is Gold and characterized by a complex
Drude dielectric function \cite{PhysRevB.92.205420} $\varepsilon_{2}%
(\omega)=1-\omega_{p}^{2}/\omega(\omega+i\gamma_{p})$ with $\omega
_{p}=1.26\times10^{16} \, rad/s$ and $\gamma_{p}=1.41\times10^{14}\,rad/s$.
The background is vacuum with $\varepsilon_{1}=1$. For simplicity, the matrix
element for the transition dipole moment is assumed to be polarized along the
radial direction of the sphere $\mathbf{d=}d \,\mathbf{\hat{r}}$ and its strength is $d=24\,D$ in this work unless otherwise specified. In section IV , we will apply the method introduced above for a QE located at the center of a plasmonic nanocavity [see Fig. 1(b)], where the permittivity $\varepsilon_{2}$ for silver is from experimental data \cite{Palik}.  The plasmonic nanocavity is composed of two silver nanorods with a gap distance $W$. Their radius and height are $R=4\,nm$ and $H=20\,nm$, respectively. The transition dipole moment is also assumed to be polarized along the axis direction.
\begin{figure}[ptbh]
\includegraphics[width=4.2cm]{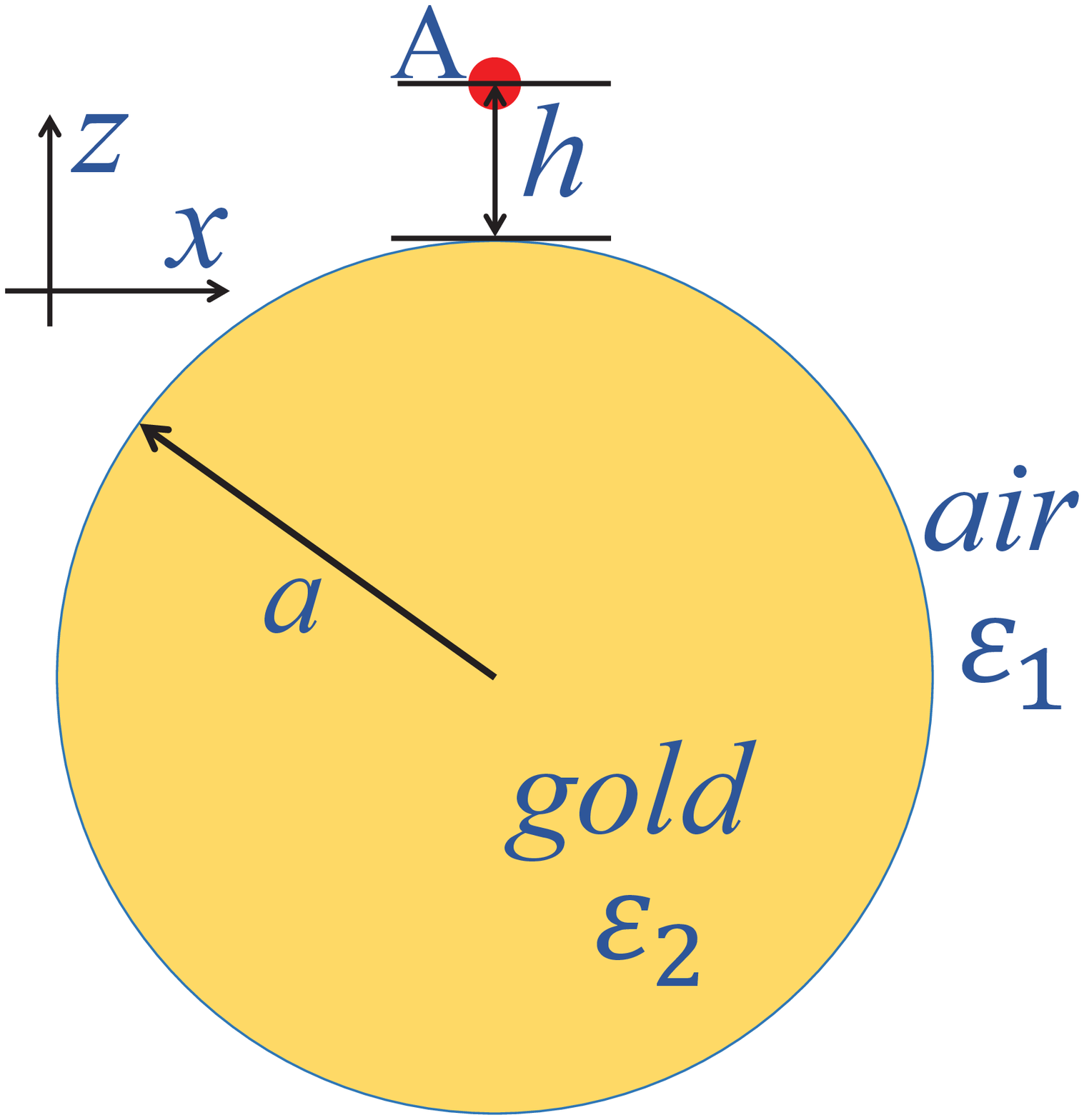}
\includegraphics[width=4.2cm]{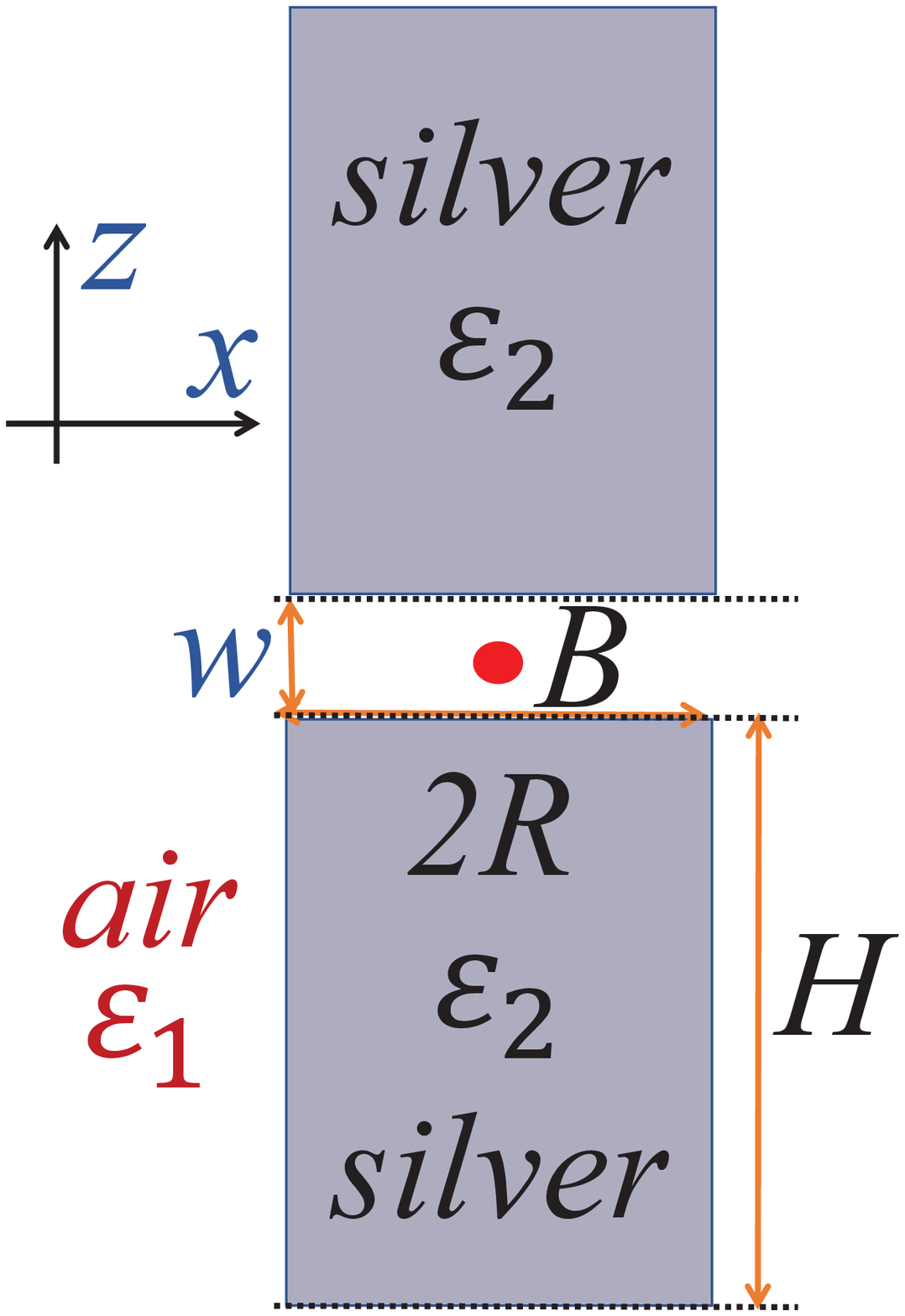}
\caption{Schematic diagrams. (a) A QE is located around a gold nanosphere with radius $a$. The distance between the emitter and the surface of sphere is $h$. (b) A QE is located at the center of two nanorods with radius $R$ and height $H$. For simplicity, the transition dipole moment for the QE is thought to be polarized along the $z$ direction. $\varepsilon_{1}$ and $\varepsilon_{2}$ are the permittivities for air and metal, respectively.}
\label{fig1}%
\end{figure}

\section{PERFROMANCES OF THE ABOVE METHODS FOR BOUND STATE AND NON-MARKOVIAN DYNAMICS}
In subsection A, we will first demonstrate the performance of the methods for  calculating the energy level shift in the positive frequency range [Eq. (\ref{SubtractiveKK})] and in the negative frequency range [(\ref{deltanefre})]. Then, the existence conditions of a bound state [Eq. (\ref{boundstatetecondition})] will be shown. In subsection B, the methods for the long-time value of the excited-state population [Eq. (\ref{zequation}) and (\ref{Ztime0})] and the non-Markovian decay dynamics [Eq. (\ref{dynamics})] are investigated. We adopt the model shown in Fig. 1(a), where the photon GF for the nanosphere can be analytically obtained \cite{Zhao:18,li1994electromagnetic,tai1994dyadic,PhysRevA.99.053844}.

\subsection{ENERGY LEVEL SHIFT AND EXISTENCE CONDITIONS OF BOUND STATE}
With the photon GF calculated by the analytical method presented in Ref. \cite{Zhao:18,li1994electromagnetic,tai1994dyadic,PhysRevA.99.053844}, the spontaneous emission rate $\Gamma(\omega)$ [Eq. (\ref{gama})] for $h=1 nm$ is shown in Fig. 2(a). The same as the result shown in Ref. \cite{PhysRevA.99.053844}, great enhancement can be observed for $\omega$ in the range of $(4-7 eV)$, which can be attributed to the localized surface plasmon resonance of the metal nanosphere.

To evaluate the correction part $\Delta c(\omega)$ for $\omega\geq0$ [Eq. (\ref{deltac})], one should calculate the real part of the coupling strength at zero frequency $\operatorname{Re}g_{rr}(0)$. Although this term can be directly obtained from the analytical GF for the nanosphere, we adopt a general method by extrapolating the photon GF near zero frequency to the static case. Different from the previous linear extrapolating method \cite{PhysRevA.99.053844}, we adopt a linear-quadratic model, in which the the real part of the coupling strength for $\omega$ near zero is written in the form of $\operatorname{Re}g_{rr}(\omega)=\alpha+\beta\omega+\gamma \omega^2$. The results for the
original data (red circle) and the extrapolating function (black solid line) are shown in Fig. 2(b). We found they agree well with $\operatorname{Re}g_{rr}(\omega)=$0.0279$-$1.1662$\times10^{-6}\omega+$0.0008$ \omega^2$. Thus, we obtain $\operatorname{Re}g_{rr}(0)=\alpha=0.0279\,meV$, which is consistent with the analytical result at extremely low frequency, for example, $\operatorname{Re}g_{rr}(\omega)=0.0279\,meV$ with $\omega=10^{-8}\,eV$.

For the integral part in Eq. (\ref{deltac}), the upper limit $+\infty$ in the integral should be replaced by some cut-off frequency $\omega c$ to perform a numerical integral. The results are shown in Fig. 2(c) with $\omega c=10\,eV$ (the black solid line) and $\omega c=200\,eV$ (the red circle). There is no observable difference and their maximum difference is less than $0.006\,meV$, which means that a small cut-off frequency ($\omega c=10\,eV$) is enough to obtain a convergent result by Eq. (\ref{deltac}).
\begin{figure}[ptbh]
\includegraphics[width=4.2cm]{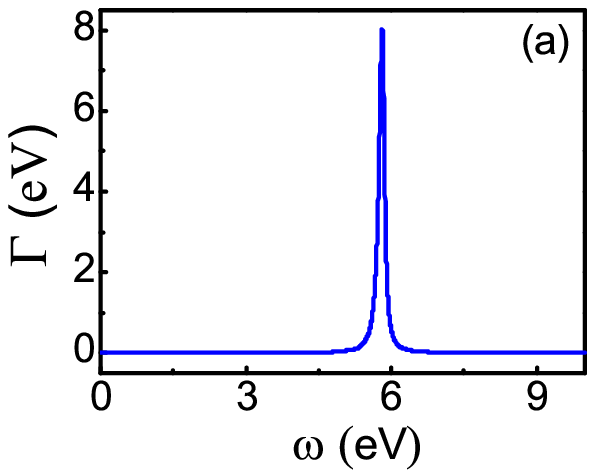}
\includegraphics[width=4.2cm]{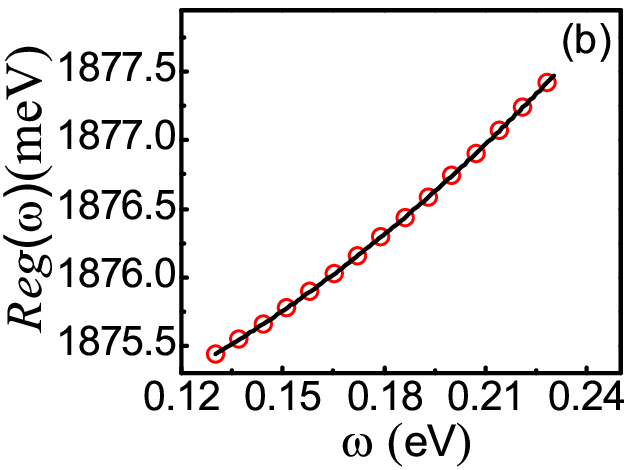}
\includegraphics[width=4.2cm]{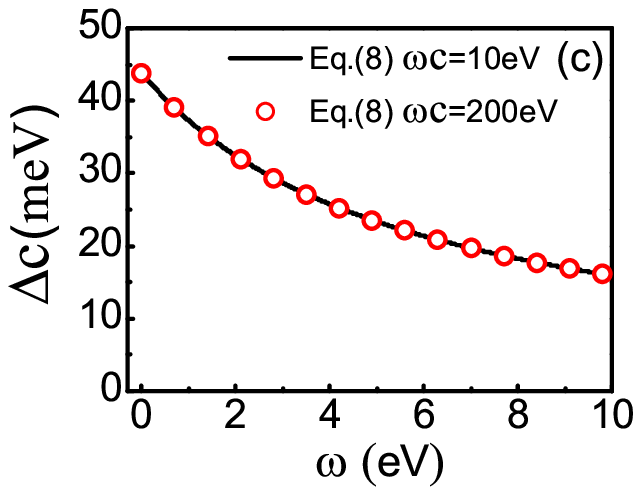}
\includegraphics[width=4.2cm]{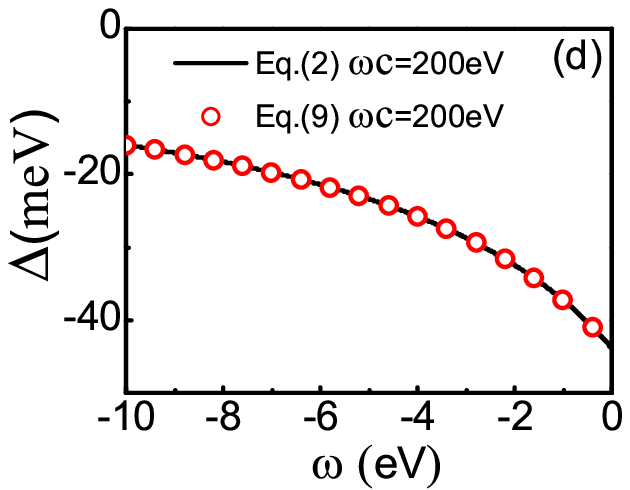}
\includegraphics[width=4.2cm]{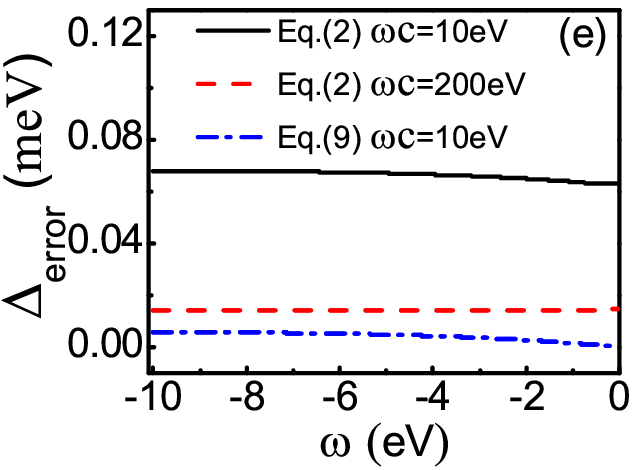}
\includegraphics[width=4.2cm]{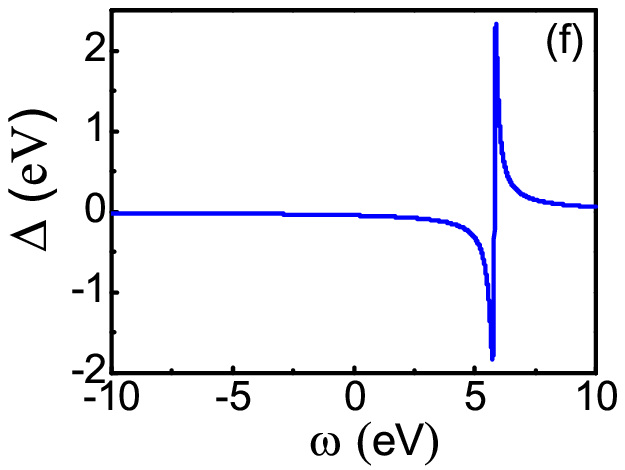}
\caption{The spontaneous emission rate and the energy level shift. (a) The spontaneous emission rate $\Gamma(\omega)$. (b) The real part of the coupling strength near zero frequency $\operatorname{Re}g_{rr}(\omega)$ (red circle) and the fitted results (black solid line). (c) The nonresonant part $\Delta c(\omega)$ for $\omega\geq0$ calculated by Eq. (\ref{deltanefre}) with $\omega c=10 \, eV$ (red circle) and $\omega c=200 \, eV$ (black solid line). (d) $\Delta(\omega)$ for $\omega<0$ calculated by Eq. (\ref{delta}) (red circle) and  by Eq. (\ref{deltanefre}) (black solid line) with $\omega c=200 \, eV$. (e)The absolute difference between $\Delta(\omega)$ obtained by Eq. (\ref{delta}) (black solid line for $\omega c=10 \, eV$ and red circle for $\omega c=200 \, eV$) or by Eq. (\ref{deltanefre}) (blue dashed line for $\omega c=10 \, eV$) and the results by Eq. (\ref{deltanefre}) with  $\omega c=200 \, eV$. (f) The energy level shift $\Delta(\omega)$.}
\label{fig2}%
\end{figure}

For a negative frequency $\omega < 0$, the energy level shift $\Delta(\omega)$ is just the negative of the correction part, i.e. $-\Delta c(-\omega)$ according to Eq. (\ref{deltanefre}) or can be calculated by the method shown in Eq. (\ref{delta}) without performing a principal value integral. The results are shown in Fig. 2(d) by Eq. (\ref{deltanefre}) (the black solid line) and  by Eq. (\ref{delta}) (red circle) with $\omega c=200 \, eV$. We find that both methods lead to almost the same results. In addition, we observe that $\Delta(\omega)$ is a monotonically decreasing function in the negative frequency range which is consistent with the theory in the previous section.

To further demonstrate the accuracy, we use the results by Eq. (\ref{deltanefre}) with a large cut-off frequency $\omega c=200 \, eV$ as a reference and show the absolute difference ($\Delta_{error}$) for the results by Eq. (\ref{delta}) with a small cut-off frequency $\omega c=10 \, eV$ (black solid line) or a large cut-off frequency $\omega c=200 \, eV$ (red dashed line) or by Eq. (\ref{deltanefre}) with $\omega c=10 \, eV$ (blue dash-dot line) in Fig. 2(e). We find that the method by Eq. (\ref{deltanefre}) converges much more quickly than the method by Eq. (\ref{delta}). Figure 2(f) demonstrates the energy level shift $\Delta(\omega)$ for our nanosphere model.
\begin{figure}[ptbh]
\includegraphics[width=4.2cm]{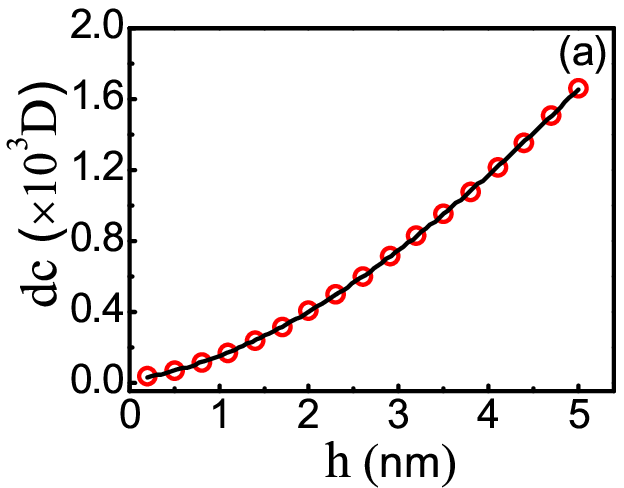}
\includegraphics[width=4.2cm]{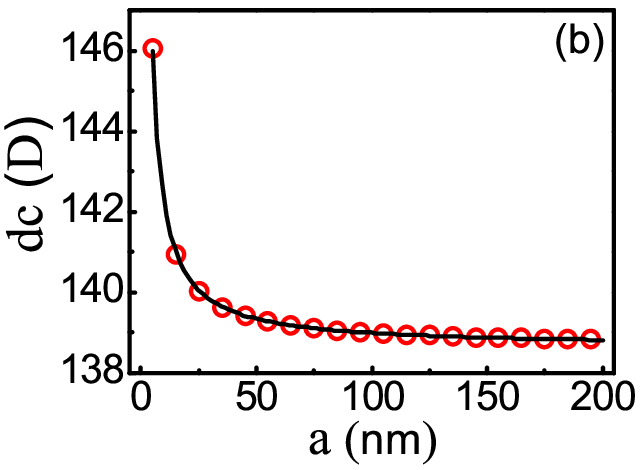}
\caption{The critical transition dipole moment $dc$ as a function of $h$ and $a$. (a) $dc$ with $a=20\,nm$. (b) $dc$ with $h=1\,nm$. The red circles are for the numerical results and the black lines are for fits.}
\label{fig3}
\end{figure}

With $\operatorname{Re}grr(0)$ obtained by the above extrapolating method, one can judge whether a bound state exists or not easily according to Eq. (\ref{boundstatetecondition}). To form a bound state, the strength for the transition dipole moment $d$ should be larger than a critical value $dc=[2\hbar\varepsilon_{0}\omega_{0}/\mathbf{\hat{r}}\cdot\operatorname{Re}\mathbf{G}(\mathbf{\mathbf{r}_{0}%
},\mathbf{r}_{0},0)\cdot\mathbf{\hat{r}}]^{1/2}$ [derived by Eq. (\ref{boundstatetecondition}) with Eq. (\ref{couplingstrg})]. Figure 3(a) shows the critical transition dipole moment $dc$ as a function of the distance $h$ between the QE and the surface of the nanosphere. The red circles are for the numerical results and the black solid line is for a cubic fit with $dc=11.177+74.563h+66.083 h^2-3.060 h^3$.  It is found that the critical dipole strength $dc$ increases sharply with the dipole-sphere distance $h$. We also vary the radius of the nanosphere $a$ with a constant dipole-surface distance $h=1 \,nm$. The results are shown in Fig. 3(b), where the red circles are for the numerical results and the black solid line is for a fit with $dc=138.868+16.943\exp(-a/4.14851)+2.472\exp(-a/31.472)$. Clearly, this shows that the critical dipole strength $dc$ decreases slowly. Thus, we can conclude that $dc$ depends heavily on the dipole-surface distance $h$ but less on the sphere radius $a$.
\subsection{LONG-TIME VALUE OF THE EXCITED-STATE POPULATION AND NON-MARKOVIAN DYNAMICS}
In this subsection, we demonstrate the performance of the two methods for calculating the amplitude of the excited state in
the long time limit $Z$ [Eq. (\ref{zequation}) and Eq. (\ref{Ztime0})] and the method for the non-Markovian decay dynamics [Eq. (\ref{dynamics})].

To evaluate $Z$ by Eq. (\ref{zequation}), we first solve the transcendental equation
$f(\omega)=\omega-\omega_{0}-\Delta(\omega)=0$ [zero of Eq. (\ref{rootequation})] to find the lowest eigenfrequency $\omega_b$. As demonstrated in Fig. 3(b), there is a critical transition dipole moment $dc=140.3\,D$ when $h=1\,nm$. We find that the root $\omega_b$ is negative when the dipole strength $d$ is beyond the critical value $dc$. For example, $\omega_b=-0.00352\,eV$ when $d=140.5\,D$. But for a slightly smaller one $d=140.1\,D$, it is positive $\omega_b=0.00346\,eV$ and there is no bound state. Figure 4(a) demonstrates the root $\omega_b$ as a function of the dipole strength $d$. It is found that the larger the dipole strength $d$ is, the lower the eigenfrequency $\omega_b$ is.

Then, we turn to the calculation of $Z$. Equation (\ref{zequation}) can be evaluated once the negative eigenfrequency $\omega_b$ is obtained. The results are shown in Fig. 4(b) (red circle). The other method to obtain $Z$ is by Eq. (\ref{Ztime0}).  We find that the results [the black solid line in Fig. 4(b)] agree well with those by Eq. (\ref{zequation}) (red circle). The inset is for the relative errors, which is the ratio of their absolute difference and their average value. It is less than $0.25\%$ for the considered transition dipole moment. It should be noted that there is no need to calculate the eigenfrequency $\omega_b$ for the method by Eq. (\ref{Ztime0}).
\begin{figure}[ptbh]
\includegraphics[width=4.2cm]{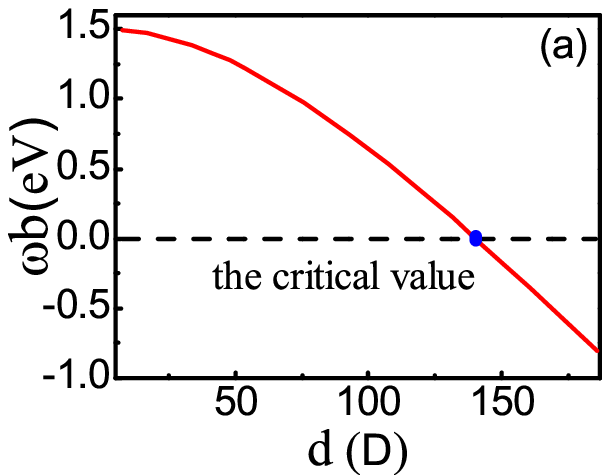}
\includegraphics[width=4.2cm]{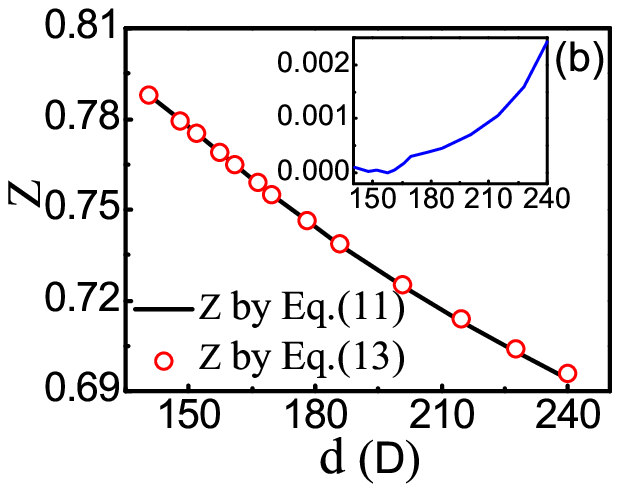}
\caption{The lowest eigenfrequency $\omega_b$ and $Z$ as a function of the dipole strength $d$. (a) $\omega_b$. (b) $Z$. The inset is for the relative difference of $Z$ obtained by Eq. (\ref{zequation}) and (\ref{Ztime0}).}
\label{fig4}%
\end{figure}

The non-Markovian dynamics [Eq. (\ref{dynamics})] are demonstrated in Fig. 5. Figure 5(a) are the results for dipole strength $d$ extremely near the critical value $dc$. The black solid line ($d=140.1\,D$), the red circle line($d=140.5\,D$) are for slightly below and above the critical value $dc=140.3\,D$, respectively. In these cases, the corresponding eigenfrequencies are positive ($\omega_b=0.00346\,eV$)and negative ($\omega_b=-0.00352\,eV$), respectively. We see that they look almost the same at short times. But for longer times (see the inset therein), a slight smaller dipole strength $d$ (black solid line) leads to an almost perfect decay when the bound state is absent. Importantly, for the case of a little larger dipole strength $d$ (red circle), the QE experiences a partial decay and becomes dissipationless after a long time. In addition, the steady-state population $P_a(\infty)$ matches well with the results $Z^2$ obtained by Eq. (\ref{zequation}) or (\ref{Ztime0}). A suppressed dissipation dynamics is observed.
\begin{figure}[ptbh]
\includegraphics[width=4.2cm]{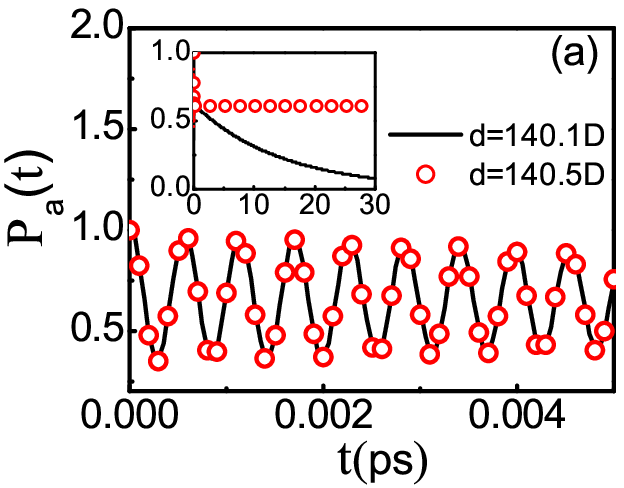}
\includegraphics[width=4.2cm]{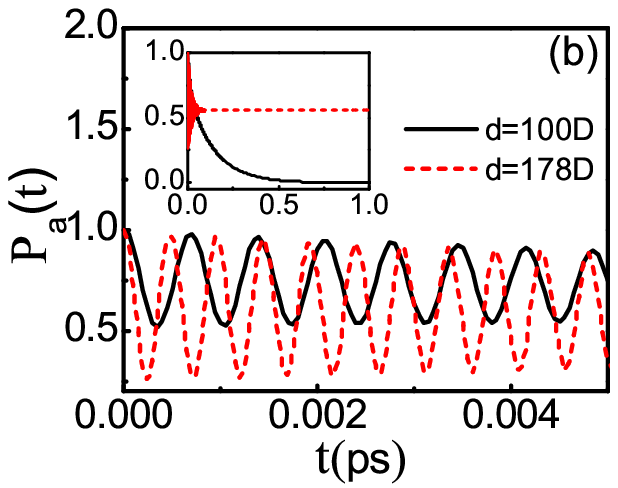}
\includegraphics[width=4.2cm]{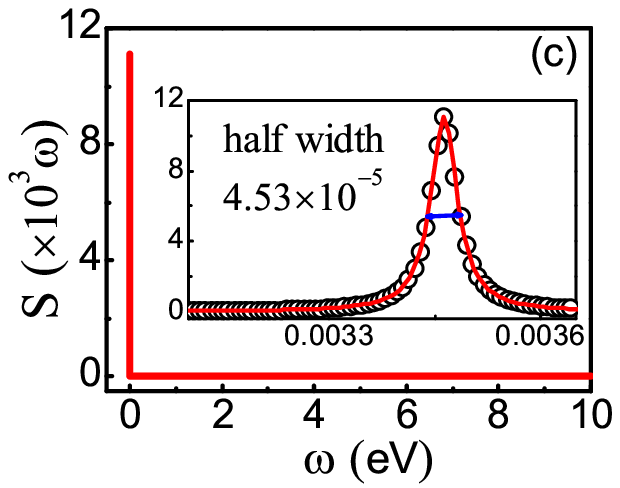}
\includegraphics[width=4.2cm]{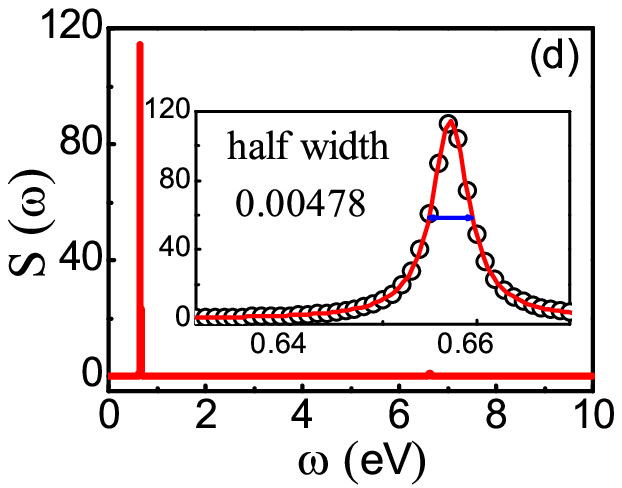}
\caption{The excited state population $P_{a}(t)=\left\vert c_{1}(t)\right\vert ^{2}$ and the evolution spectrum $S(\omega)$ in the positive frequency range. (a) $P_{a}(t)$ for $d$ around the critical value $dc=140.3\,D$. The red circles are for a little larger transition dipole moment $d=140.5\,D$ and the black solid line is for a little smaller dipole moment $d=140.1\,D$. (b) The same as (a) but for a much larger ($d=178\,D$ red dash line) and a much smaller ($d=100\,D$ black solid line) transition dipole moment. (c) $S(\omega)$ for $d=140.1\,D$. (d)$S(\omega)$ for $d=100\,D$. The insets in (c) and (d) show the Lorentz fits with half widths $w1=0.0453\,meV$ and $w2=4.78\,meV$, respectively.}
\label{fig5}%
\end{figure}

To further demonstrate this, we choose another two different values of the dipole strength $d$. One is much larger than the critical value $dc$ and the other is much smaller. The results are shown in Fig. 5(b). The red dashed line ($d=178\,D$ leading to a negative root $\omega_b=-0.65967\,eV$) and the black solid line ($d=100\,D$ leading to a positive root $\omega_b=0.65725\,eV$) demonstrate a totally different decay characteristics.
Different from the case shown in Fig. 5(a),
the short time behaviors of the decay dynamics are different for these two values of dipole strength $d$. In addition, the time needed for the complete decay is much shorter for $d=100\,D$ than that for $d=140.1\,D$ [see the black solid lines in the insets of Fig. 5(a) and 5(b)]. This can be understood by checking the evolution spectrum $S(\omega)$ in the positive frequency domain (see Fig. 5(c) and 5(d) for $d=140.1\,D$ and $d=100\,D$, respectively). We find that $S(\omega)$ around the peak value can be modeled by a perfect Lorentz function in both cases [see the insets in Fig. 5(c) and (d), where the red circles are for numerical results and the black solid lines are for their Lorentz fits]. The half widths, which represent the spontaneous emission rates, are $w1=0.0453\,meV$ and $w2=4.78\,meV$ for the above two different dipole moments, which is approximately two orders of magnitude difference.
\begin{figure}[ptbh]
\includegraphics[width=8.4cm]{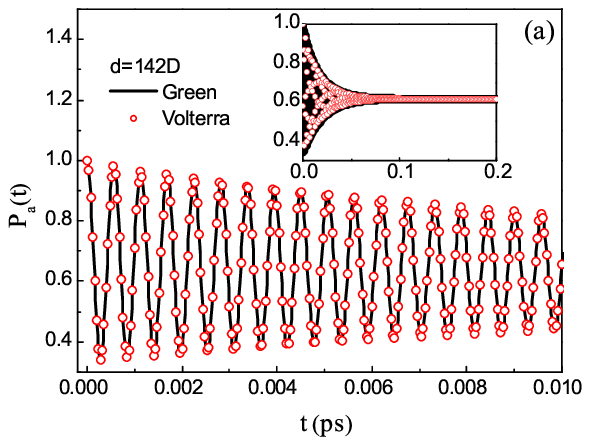}
\includegraphics[width=8.4cm]{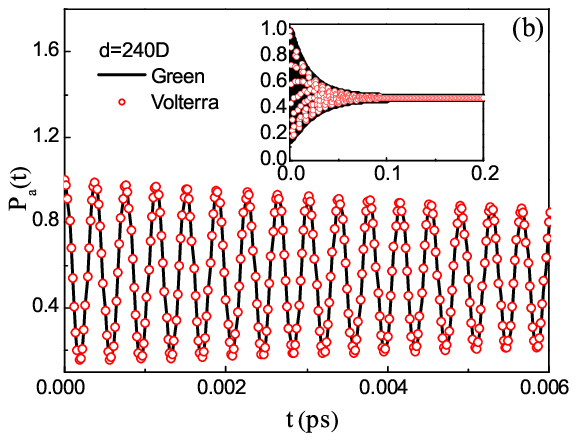}
\caption{Performance of the method [Eq. (\ref{dynamics})] for the excited state population $P_{a}(t)=\left\vert c_{1}(t)\right\vert ^{2}$ when a bound state exists with $d$ around and away from the critical value $dc=140.3\,D$. (a) $d=142\,D$. (b) $d=240\,D$. The black solid lines are the results by the method based on the Green¡¯s function expression for the evolution operator [Eq. (\ref{dynamics})]. The red circles are for the results by solving the non-Markovian Schr\"{o}dinger equation (Volterra integral equation of the second kind) in the time domain [Eq. (3) in Ref. \cite{PhysRevA.99.053844}, i.e. $c_{1}(t)e^{i\omega_0 t}=c_{1}(0)-\int_{0}^{t}B(t-s)c_{1}(s)ds$ with $B(t-s)=\int_{0}^{+\infty}d\omega \operatorname{Im}g(\omega)\int
_{0}^{t-s}e^{-i(\omega-\omega_0)u}du$]. The results agree well, which means that Eq. (\ref{dynamics}) is able to obtain the exact non-Markovian dynamic when a bound state presents.}
\label{fig6}%
\end{figure}

As demonstrated in Ref. \cite{PhysRevA.99.053844}, the decay dynamics can also be investigated by solving the non-Markovian  Schr\"{o}dinger equation in the time domain, which is equivalent to the method adopted in Ref. \cite{PhysRevB.95.161408}. Although this method is time consuming, we compare it with the resolvent operator method [Eq. (\ref{dynamics})] when a bound state appears. Figure 6(a) and 6(b) are the results for $d=142\,D$ and $d=240\,D$, respectively. The lowest eigenfrequencies are $\omega_b=-0.0278\,eV$ and $\omega_b=-1.8209\,eV$ respectively, which are around and far away from zero. We find that the results for the excited state population $P_{a}(t)$ by both methods agree well in both cases. The insets are for the long-time results. It should be noted that the resolvent operator method requires much less computation time than the method by solving the Schr\"{o}dinger equation in the time domain. This clearly demonstrates that Eq. (\ref{dynamics}) can be used to efficiently and accurately evaluate the non-Markovian decay dynamics when a bound state presents.

The main results of the above two sections can be summarized as follows. We have demonstrated that the real part of the coupling strength at zero frequency $Reg_{rr}(0)$ can be obtained by the extrapolating method. For the energy level shift, we have proven that the method by Eq. (\ref{deltanefre}) is more efficient than the method by Eq. (\ref{delta}). Thus, the exact energy level shift can be obtained efficiently by Eq. (\ref{SubtractiveKK}) for positive frequency and by Eq. (\ref{deltanefre}) for negative frequency. Besides, one can quickly judge whether a bound state exists or not by Eq. (\ref{boundstatetecondition}) with $Reg_{rr}(0)$ obtained. We have proven that results for the non-Markovian decay dynamics by Eq. (\ref{dynamics}) or by solving the non-Markovian Schr\"{o}dinger equation (see Ref. \cite{PhysRevA.99.053844,PhysRevB.95.161408}) are the same. Thus, the non-Markovian decay dynamics can be exactly obtained by Eq. (\ref{dynamics}), in which the parameter $Z$ can be either calculated by Eq. (\ref{zequation}) or (\ref{Ztime0}). In spite of this, one need not to evaluate the lowest eigenfrequency $\omega_b$ [negative zero for the transcendental Eq. (\ref{rootequation})] for the method by Eq. (\ref{Ztime0}) compared to the method by Eq. (\ref{zequation}). We will use these methods to investigate the existence condition of a bound state and the non-Markovian decay dynamics when a QE is located in a gap plasmonic nanocavity.
\section{EXISTENCE CONDITIONS OF A BOUND STATE AND NON-MARKOVIAN DYNAMICS OF A QE IN A PLASMONIC NANOCAVITY}

In this section, we adopt the model shown in Fig. 1(b). Different from the previous section where the Drude model for the permittivity of metal over the whole frequency range is assumed, the permittivity $\varepsilon_{2}$ for silver is from experimental data \cite{Palik}. The photon GF for the nanocavity  is numerically obtained by COMSOL MULTIPHYSICS software with the method presented in Ref. \cite{yunjinwulixuebao,Zhao:18,PhysRevA.99.053844}.  We first vary the geometric parameters of the nanocavity, such as the height $H$ and the radius $R$ of the nanorod, to find the critical dipole moment $dc$ for the system to form a bound state according to Eq. (\ref{boundstatetecondition}). Then, we investigate the long-time value of the excited-state population by Eq. (\ref{zequation}) and (\ref{Ztime0}). At the end, the non-Markovian dynamics by solving Eq. (\ref{dynamics}) are demonstrated. For simplicity, the transition frequency for the QE is also set to be $1.5\,eV$.

\begin{figure}[ptbh]
\includegraphics[width=4.2cm]{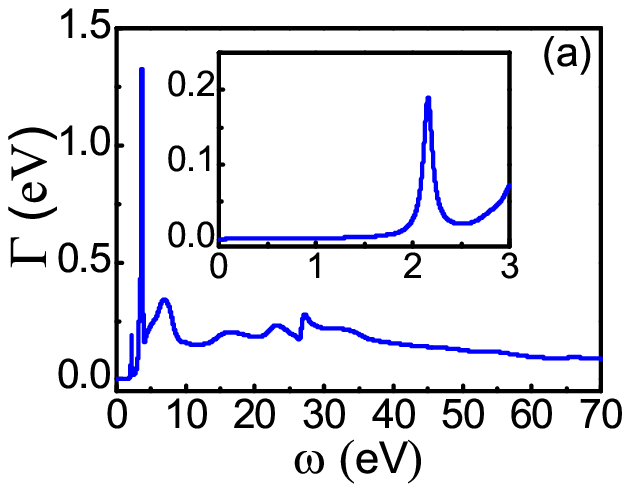}
\includegraphics[width=4.2cm]{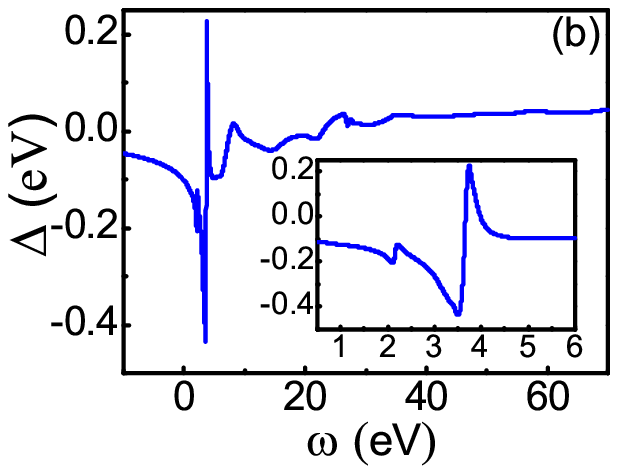}
\includegraphics[width=4.2cm]{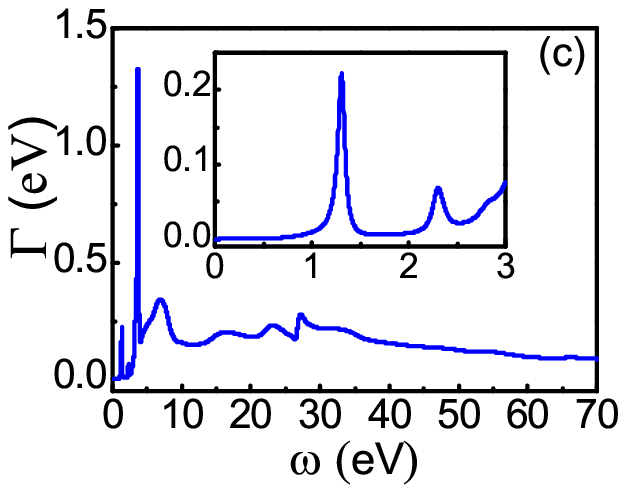}
\includegraphics[width=4.2cm]{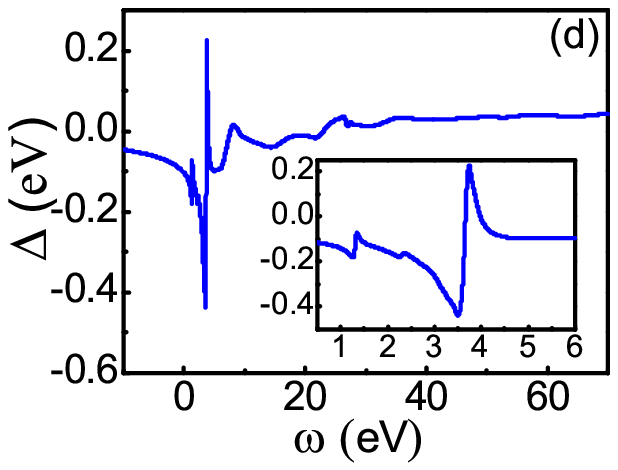}
\caption{The spontaneous emission rate $\Gamma(\omega)$ and the energy level shift $\Delta(\omega)$ for the model shown in Fig. 1(b). (a) and (b) are for $H=20\,nm$. (c) and (d) are for $H=50\,nm$. Both $\Gamma(\omega)$ and $\Delta(\omega)$ are relative large at high frequency. The insets show the results for frequency around the lowest plasmonic mode. Here, $d=24\,D$.}
\label{fig7}%
\end{figure}

The spontaneous emission rate $\Gamma(\omega)$ and the energy level shift $\Delta(\omega)$ are shown in Fig. 7 for two different nanocavities with $H=20nm$ and $H=50nm$. Different from the above section where both $\Gamma(\omega)$ and $\Delta(\omega)$ take great value only in a much narrow frequency range, they are relative large over a wide frequency range. This is because the permittivity for metal is from experimental data, which is beyond the Drude model and takes into account the realistic response of metal at high frequency. From the insets in Fig. 7, we observe that the lowest plasmonic mode contributes much to the spontaneous emission rate $\Gamma(\omega)$ and the energy level shift $\Delta(\omega)$,  in which the peak position and peak value depend heavily on the rod height $H$.

Although $\Gamma(\omega)$ and $\Delta(\omega)$ look very different for the above different nanocavities, the critical dipole moment $dc$ for the system to form a bound state is nearly independent of the radius $R$ and height $H$ of the nanorod. This can be clearly seen from the second column of Table \ref{tab:table1}. $Z1$ and $Z2$ are the amplitude for the excited state in the steady limit obtained by Eq. (\ref{zequation}) and (\ref{Ztime0}), respectively. Their differences are very small (see Table \ref{tab:table1}), which imply that Eq. (\ref{zequation}) and (\ref{Ztime0}) are able to calculate $Z$. It should be noted that there is no need to calculate the eigenfrequency $\omega_b$ by Eq. (\ref{Ztime0}). In addition, we observe that $Z$ is nearly independent of the sizes of the nanorod and the transition dipole moment $d$. But for the negative eigenfrequency $\omega_b$, it depends heavily on them, especially on the dipole strength $d$.
\begin{table*}
\caption{\label{tab:table1}The results of $dc$, $\omega_b$ and $Z$ with different heights $H$ and radius $R$ of the nanorod. $dc$ is the critical strength for the dipole moment obtained by Eq. (\ref{boundstatetecondition}). $\omega_b$ is the negative eigenfrequency for $f(\omega)=\omega-\omega_{0}-\Delta(\omega)=0$ [zero of Eq. (\ref{rootequation})]. $Z1$ and $Z2$ are the amplitude for the excited state in the steady limit obtained by Eq. (\ref{zequation}) and (\ref{Ztime0}), respectively. Their differences are very small. We observe that $dc$ and $Z$ are nearly independent of the height $H$ and radius $R$ of the nanorod. But $\omega_b$  depends heavily on them, especially on the transition dipole moment $d$. }
\begin{ruledtabular}
\begin{tabular}{ccccccccccc}
 &&\multicolumn{3}{c}{$d=dc+0.01\,D$}&\multicolumn{3}{c}{$d=100\,D$}&\multicolumn{3}{c}{$d=120D$}\\ \cline{3-10}
 $(H,R)\,(nm)$&$dc \,(D)$&$\omega_b\,(eV)$&$Z1$&$Z2$&$\omega_b\,(eV)$&$Z1$&$Z2$&$\omega_b\,(eV)$&$Z1$&$Z2$\\ \hline
 $(20,4)$&$92.1740$&$-0.0008$&$0.8031$&$0.8072$&$-0.2139$&$0.8007$&$0.8048$&$-0.7816$&$0.7772$&$0.7818$\\
 $(20,8)$&$90.6460$&$-0.0009$&$0.8039$&$0.8093$&$-0.2603$&$0.7982$&$0.8039$&$-0.8407$&$0.7734$&$0.7798$\\
 $(20,20)$&$89.7844$&$-0.0009$&$0.8001$&$0.8053$&$-0.2860$&$0.7936$&$0.7992$&$-0.8711$&$0.7692$&$0.7756$\\
 $(50,4)$&$91.3852$&$-0.0008$&$0.7809$&$0.7880$&$-0.2332$&$0.7846$&$0.7914$&$-0.7949$&$0.7666$&$0.7738$\\
$(80,4)$&$90.9872$&$-0.0008$&$0.7662$&$0.7703$&$-0.2432$&$0.7788$&$0.7829$&$-0.8042$&$0.7652$&$0.7701$\\
$(100,4)$&$90.8656$&$-0.0008$&$0.7594$&$0.7636$&$-0.2464$&$0.7777$&$0.7816$&$-0.8077$&$0.7654$&$0.7699$\\
\end{tabular}
\end{ruledtabular}
\end{table*}

The decay dynamics are shown in Fig. 8. Here, the height and the radius for the nanorod are $H=20\,nm$ and $R=4\,nm$, respectively. The critical dipole strength to form a bound state is $dc=92.174\,D$ for a transition frequency $\omega_0=1.5\,eV$. Figure 8(a) demonstrates the excited state population $P_{a}(t)$ for the dipole strength $d$ just below and above its critical value $dc$. At the very beginning, there is little difference, which is similar to the phenomenon shown in Fig. 5(a). The difference increases with time. For a little smaller dipole strength $d=90\,D$ , we observe a complete decay where the excited state population $P_{a}(t)$ tends to zero [see the black solid line in the inset of Fig. 8(a)]. But for a slightly larger dipole strength $d=94\,D$, the excited state population $P_{a}(t)$ is partially preserved and tends to $0.64$ as $t\rightarrow\infty$.

We also consider another two different dipole moments, which are much smaller or larger than the critical value $dc$. In this case, the excited state population $P_{a}(t)$ for the above two different dipole strengths behaves differently at the beginning. This implies that the decay dynamics is much affected by the dipole strength at the beginning. For the long time performance, a complete decay and a partial limited decay can be observed depending on the dipole moment [see the insets in Fig. 8(b)]. For $d=130\,D$, $P_{a}(\infty)=0.593$, which is around $Z^2=0.604$ for $d=120\,D$ (see the last two columns in Table \ref{tab:table1}). This is also consistent with the argument that the dipole strength $d$ takes little effect on the excited state population in the long time limit.

The above phenomena can also exist for another plasmonic nanocavity composed of nanorods with different geometric parameter. Figure 8(c) and 8(d) demonstrate the results for a higher nanorod $H=80\,nm$ with the same radius $R=4\,nm$. We find that the results shown in Fig. 8(c) looks similar to that in Fig. 8(a). Differently, the time for the system to reach its steady state is much smaller for Fig. 8(d) than that in Fig. 8(b).
\begin{figure}[ptbh]
\includegraphics[width=4.2cm]{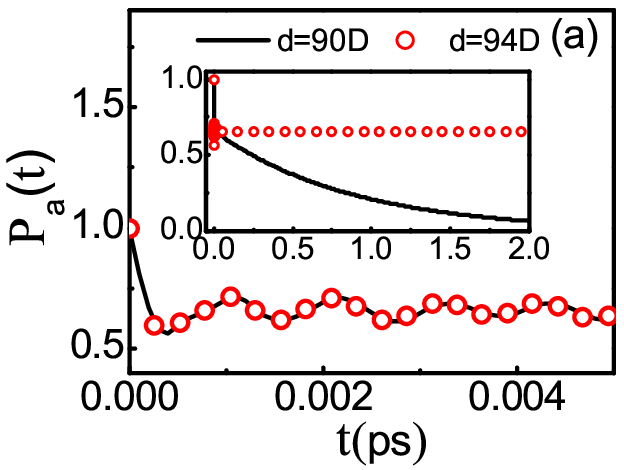}
\includegraphics[width=4.2cm]{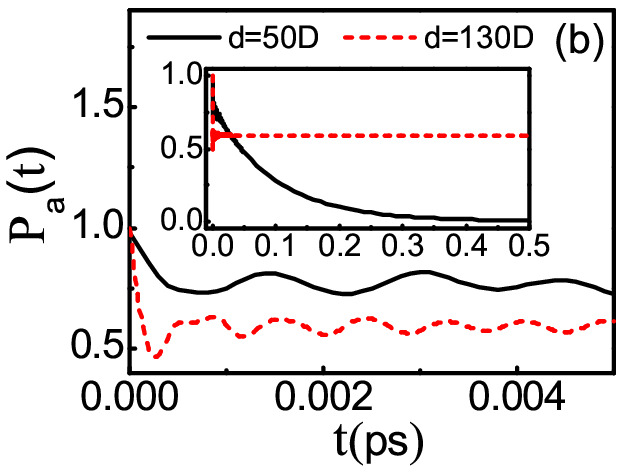}
\includegraphics[width=4.2cm]{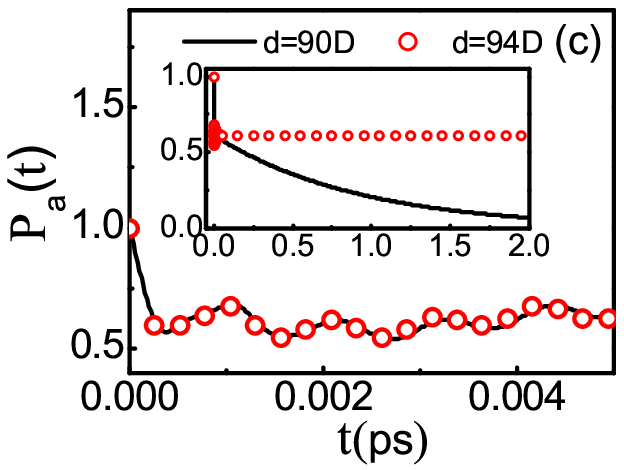}
\includegraphics[width=4.2cm]{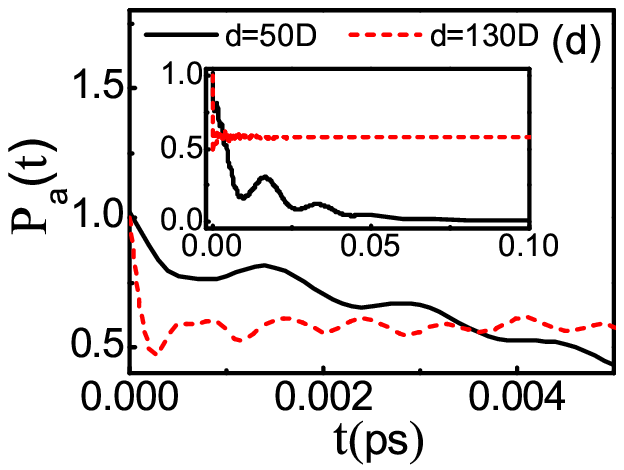}
\caption{The excited state population $P_{a}(t)=\left\vert c_{1}(t)\right\vert ^{2}$. (a) $d$ is  a little larger ($d=94\,D$ red circles) and smaller ($d=90\,D$ black solid line) than the critical value $dc$. (b) $d$ is  much larger ($d=130\,D$ red dash line) and much smaller ($d=50\,D$ black solid line) than the critical value $dc$. Here, $dc=92.1740\,D$ for $H=20\,nm$. (c) and (d) are similar to (a) and (b) except for a much higher rod $H=80\,nm$. In this case, the critical transition dipole moment is $dc=90.9872\,D$. Here, the radius of the nanorod is $R=4\,nm$.}
\label{fig8}%
\end{figure}

\section{CONCLUSIONS}
In summary, we have presented an efficient numerical method [Eq. (\ref{boundstatetecondition})] to determine the existence or absence of a bound state for a QE coupled to surface plasmon polaritons without calculating the time evolution of the system. We have found that the critical dipole strength $dc$ is heavily dependent on the dipole-sphere distance $h$ but much less dependent on the sphere-radius $a$ for the QE located around a nanosphere. For the QE at the center of a plasmonic nanocavity composed of two nanorods, we have found that $dc$ is nearly independent of the radius and height of the rod. For a bound state presented, we have proposed two different methods to determine the long-time value for the excited state population $P_a(\infty)=Z^2$ [Eq. (\ref{zequation}) and (\ref{Ztime0})]. To evaluate Eq. (\ref{zequation}), we have proposed a new formalism to calculate the energy level shift for a negative frequency [Eq. (\ref{deltanefre})], which is more efficient than the method by Eq (\ref{delta}). This is helpful to determine the exact eigenfrequency $\omega_b$ [zero for Eq. (\ref{rootequation})] for the bound state, which is an important parameter to evaluate $Z$ by Eq. (\ref{zequation}). For a QE located around a nanosphere or in a plasmonic nanocavity, we have shown that both methods [Eq. (\ref{zequation}) and (\ref{Ztime0})] lead to the same results. In addition, we have found that $Z$ is less affected by the dipole strength $d$ with $d\geq dc$, but the lowest eigenfrequency $\omega_b$ is much dependent on it.

By comparing with the time-domain method via solving the non-Markovian Schr\"{o}dinger equation, we have shown that the non-Markovian decay dynamics in the presence of a bound state can be exactly and efficiently obtained by the method based on the Green's function expression for the evolution operator. It is found that the excited state population $P_a(\infty)$ matches well
with $Z^2$ obtained by Eq. (\ref{zequation}) or (\ref{Ztime0}). In addition, we have found that $P_a(\infty)$ is nearly independent of the sizes of the nanorod and the transition dipole strength $d$ when $d$ is larger than the critical value $dc$, which is consistent with the prediction by Eq. (\ref{zequation}) and (\ref{Ztime0}).

\begin{acknowledgments}
This work was financially supported by the National Natural Science Foundation
of China (Grants No. 11564013, 11464014, 11464013) and
Hunan Provincial Innovation Foundation For Postgraduate (Grants No.CX2018B706).
\end{acknowledgments}

\bibliography{20190506}

\end{document}